\documentclass[prd,floats,superscriptaddress,nofootinbib,preprintnumbers]{revtex4}%

\usepackage{lipsum, babel}

\usepackage{amssymb}
\usepackage{amsmath}

\usepackage{sans}

\makeatletter\AtBeginDocument{\let\@elt\relax}\makeatother

\usepackage{graphicx}
\usepackage{graphics}
\usepackage{dcolumn}
\usepackage{color}
\usepackage{rotate}
\usepackage{fancyhdr} 
\usepackage{hyperref}
\usepackage{indentfirst}
\usepackage{booktabs} 
\usepackage{fontawesome} 
\usepackage{physics} 
\usepackage{siunitx} 
\DeclareSIUnit \parsec {pc}
\DeclareSIUnit \year {yr}
\usepackage{umoline}

\def\be{\begin{eqnarray}}
\def\ee{\end{eqnarray}}
\def\ba{\begin{eqnarray}}
\def\ea{\end{eqnarray}}
\def\no{\nonumber}

\definecolor{darkred}{rgb}{.743,0,0}

\begin{document}
\title{Unraveling the bounce: a real time perspective on tunneling}

\author{Kfir Blum}\email{kfir.blum@weizmann.ac.il}
\affiliation{Weizmann Institute of Science, Rehovot, Israel} 
\author{Omri Rosner}\email{omri.rosner@weizmann.ac.il}
\affiliation{Weizmann Institute of Science, Rehovot, Israel} 

\begin{abstract} 
We study tunneling in one-dimensional quantum mechanics using the path integral in real time, where solutions of the classical equation of motion live in the complex plane. Analyzing solutions with small (complex) energy, relevant for constructing the wave function after a long time, we unravel the analytic structure of the action, and show explicitly how the imaginary time bounce arises as a parameterization of the lowest order term in the energy expansion. The real time calculation naturally extends to describe the wave function in the free region of the potential, reproducing the usual WKB approximation. The extension of our analysis to the semiclassical correction due to fluctuations on the saddle is left for future work.
\end{abstract}

\maketitle


\section{Introduction}\label{s:int}
Vacuum decay through barrier penetration is typically considered in terms of the survival probability associated with a state that is initially localized in a false vacuum (FV) region of the potential~\cite{Coleman:1977py,Callan:1977pt,Schulman:1981vu,Kleinert:2004ev,Andreassen:2016cvx}: $P_{FV}(t)=\int_{FV}dx|\psi(x,t)|^2$. 
Here $|\psi(t)\rangle$ is a Schr\"odinger state and it is assumed that $P_{\rm FV}(0)\approx1$. At late time after transients fade out, but not so late that returning current matters~\cite{Andreassen:2016cvx}, one finds exponential decay $P_{FV}(t)\propto e^{-\Gamma t}$. 
Information about the tunneling process is contained in the wave function
$\psi(y,t)=\int dx\,\psi_0(x)K(t;y,x)$, 
with $\psi_0(x)=\langle x|\psi(0)\rangle$ and the propagator
$K(t;y,x)=\langle y|e^{-iHt}|x\rangle$. 
In this paper we focus on the path integral representation of the propagator,
\be K(t;y,x)&=&\underset{\scriptsize{\begin{array}{c}z(0)=x\\ z(t)=y\end{array}}}{\int  Dz} e^{iS[z]},
\;\;\;\;\;\;S[z]\,=\,\int_0^tdt'\left(\frac{1}{2}\dot z^2-V(z)\right).\ee
We will restrict our attention to contributions to the propagator that become relevant for the typical initial data $\psi_0(x)$ of the tunneling problem.

One learns to calculate $\Gamma$ by considering the FV-to-FV propagator continued to imaginary time, $K(t;0,0)\to K(-i\tau;0,0)=K_E(\tau;0,0)$ (subscript $_E$ for Euclidean)~\cite{Coleman:1977py,Callan:1977pt,Schulman:1981vu,Kleinert:2004ev}. In a nutshell, in the limit of large $\tau$, the path integral for $K_E(\tau;0,0)$ receives a contribution from a real-valued saddle point function $r(\tau)$ that solves the equation of motion (EOM) $\frac{d^2r}{d\tau^2}=V'(r)$ with boundary conditions $r(\pm\infty)\to0$. The Euclidean action of this solution is twice $S_E[r]=\int_0^b dr\sqrt{2V(r)}$, where $b$ is the classical turning point of the potential (see Fig.~\ref{fig:V}). Including corrections due to fluctuations around $r(\tau)$, one finds $\Gamma\propto e^{-2S_E}$.

Here, instead of going to imaginary time, we consider the path integral in real time. Our goal is to investigate how the usual imaginary time result arises from the real time path integral. The analysis allows us, at least in principle, to calculate finite time corrections to the Euclidean result. It also allows us to extend the calculation of the large-$t$ propagator to arbitrary points outside the barrier, $y>b$, a calculation that is omitted in the imaginary time formalism. Another way to state our exercise is as a real time path integral derivation of the usual WKB tunneling wave function.

The semiclassical approximation expands around classical paths ${z_c}(t)$ that satisfy the real time EOM and boundary conditions:
\be\label{eq:zbar}{\ddot z_c}+V'&=&0,\\
{z_c}(0)&=&x,\;\;\;{z_c}(t)=y.\ee
Solutions of Eq.~(\ref{eq:zbar}) have a constant of motion, the energy $\epsilon$:
\be\label{eq:eps} \epsilon&=&\frac{1}{2}{\dot z_c}^2+V.\ee
The semiclassical approximation for the propagator is 
\be \underset{\scriptsize{\begin{array}{c}z(0)=x\\ z(t)=y\end{array}}}{\int  Dz} e^{iS[z]}&=&e^{iS[{z_c}]}\mathcal{A}.\ee
The prefactor $\mathcal{A}$ contains the path integral over fluctuations on the classical path. If more then one classical path exists, one needs to sum $\sum_je^{iS[{z_c}_j]}\mathcal{A}_j$.

We are interested in the action of small-$\epsilon$ solutions of the EOM, with $|\epsilon|$ much smaller than the potential barrier. We call these tunneling solutions. 
Tunneling solutions live in the complex $z$ plane, and have complex $\epsilon$.  
Extending the path integral to complex paths requires a restriction to guarantee that the dimensionality of the path space is conserved. Picard-Lefschetz theory provides this machinery by arranging the path integral as a sum over complex saddles, where the path integration in the vicinity of each saddle is constrained by a flow equation~\cite{Witten:2010zr,Witten:2010cx,Basar:2013eka,Cristoforetti:2013wha,Tanizaki:2014xba,Cherman:2014sba,Behtash:2015loa,Alexandru:2016gsd,Hertzberg:2019wgx,Ai:2019fri,Mou:2019tck,Mou:2019gyl,Nishimura:2023dky}\footnote{Complex solutions of the classical EOM were related by~\cite{Bender:2008fr} to quantum phenomena like energy quantization. The complexified path integral method makes it natural to attribute these relations to the usual emergence of classical mechanics from extremal action paths~\cite{Turok:2013dfa}.}. 

Ref.~\cite{Tanizaki:2014xba} provided a useful summary of the formalism, and we refer the reader there for details. 
The first stage of the calculation, which yields the leading semiclassical factor $e^{iS[{z_c}]}$, is the usual task of finding the saddle point paths and calculating their action. The only change compared to the real path integral is that the saddle now extends to complex configurations. More involved features of the theory arise in the calculation of the integral over fluctuations. However, we leave this interesting and important part of the calculation out of the current paper. We plan to report the analysis of the fluctuation integral in subsequent work. Leaving out the fluctuation analysis allows us to put the spotlight on the leading semiclassical object, the real time parallel of Coleman's bounce\footnote{To be precise, we will focus on a solution that connects the FV to the free region, so the Euclidean parallel is the instanton, or half of a bounce.}. 

Many analyses of tunneling in the literature refer to real time path integrals~\cite{Turok:2013dfa,Tanizaki:2014xba,Cherman:2014sba,Bramberger:2016yog,Hertzberg:2019wgx,Ai:2019fri,Mou:2019gyl,Nishimura:2023dky}\footnote{See also~\cite{Levkov:2004ij} and~\cite{Braden:2018tky} for related discussions.}. Our analysis differs from previous literature in that we do not explore the analytic continuation of the amplitude w.r.t. the time variable. Instead, we stubbornly stick to real time and unravel the complex bounce, exploring how the result of the calculation maps to the imaginary time result by means of an explicit expansion in powers of (complex) path energy; or (almost) equivalently, inverse powers of real physical time. The analysis reveals how the basic WKB Lorentzian-Euclidean factorization, highlighted in~\cite{Bramberger:2016yog}, arises in the limit of large time.

\section{$K(t;y,0)$: propagator starting at the false minimum and ending in the free region}
\begin{figure}[h]
      \includegraphics[scale=0.5]{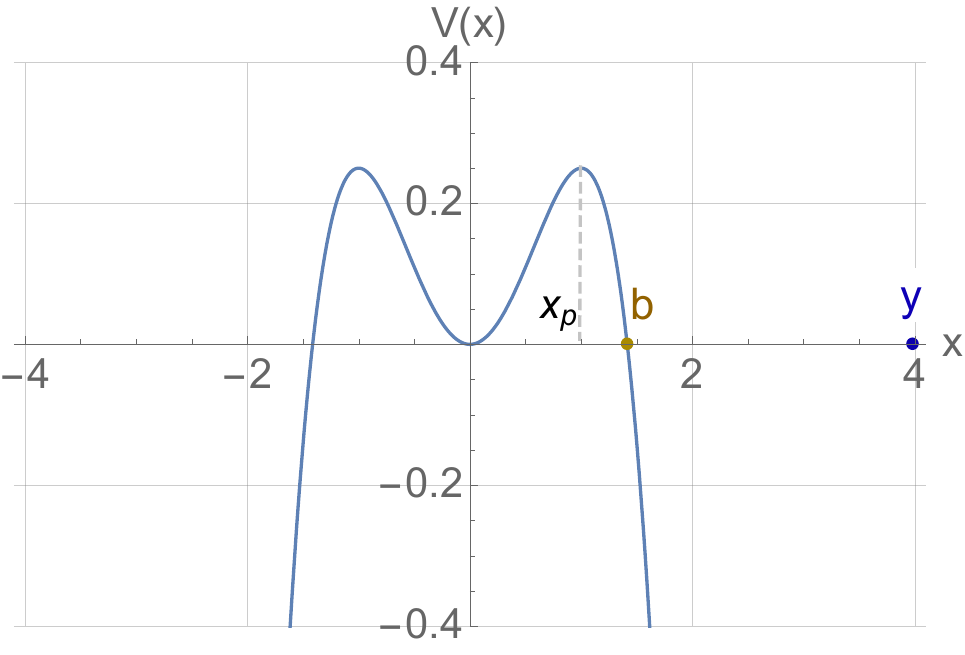} 
 \caption{Potential evaluated on the real axis, for $n=4$ (see Eq.~(\ref{eq:Vzn})). 
 }
 \label{fig:V}
\end{figure}

We consider potentials that can be approximated by a parabola at the FV minimum. A suitable example is
\be\label{eq:Vzn} V(z)&=&\frac{1}{2}z^2-\frac{1}{n}z^n,\ee
assuming $n\geq3$. The peak of this potential is at $x_p=1$, and we have $V(x_p)=\frac{1}{2}-\frac{1}{n}\sim\mathcal{O}(1)$. 
This is sufficiently general for us to use as concrete example\footnote{Eq.~(\ref{eq:Vzn}) arises from the action $S=\int dt\left(\frac{1}{2}\dot z^2-\frac{m^2}{2}z^2+\frac{1}{n}l^{2-n} m^2 z^n\right)$, with mass and length parameters $m$ and $l$. 
Defining $t\to mt$ and $z\to l z$, we have $S=(ml)^2\int dt\left(\frac{1}{2}\dot z^2-\frac{1}{2}z^2+\frac{1}{n}z^n\right)$.}. It will become clear that the most important behaviour at small $\epsilon$ is independent of the details of $V(z)$. A potential with $n=4$ is shown in Fig.~\ref{fig:V}.

Coleman's original derivation~\cite{Coleman:1977py} focused on $K(t;0,0)$, a restricted version of the FV-to-FV propagator. We will consider the slightly different calculation of $K(t;y,0)$, that is, starting point $x\approx0$ near the FV minimum, but end point $y$ in the free region. These calculations encode similar physics.

\subsection{Choosing the correct saddle}\label{ss:sad}
Before entering any details of the calculation, the first point to note is that for real potentials (in our case, potentials that are polynomials with real coefficients), any solution $z_c(t)$ of the complexified EOM with real boundary conditions $x,y$ always comes with a complex conjugate partner solution $z_c^*(t)$. If the energy corresponding with $z_c$ is $\epsilon$, the energy of $z^*_c$ is $\epsilon^*$. 
Similarly, the action is also related by complex conjugation: $S[z^*_c]=S^*[z_c]$. Thus $iS[z_c^*]=-(iS[z_c])^*$. Now, when the dust settles, our analysis will (reassuringly) re-discover the usual WKB result that $iS[z_c]=-S_E+iS_{\rm free}$, where $S_E$ and $S_{\rm free}$ are real functions of $x$ and $y$, respectively. It follows that $S_E[z_c^*]=-S_E[z_c]$: namely, one of the complex conjugate pair of classical paths has a positive Euclidean action, and the other, negative. Throughout this paper we focus on the solution with $S_E>0$, that we simply denote by $z_c(t)$. Of the complex conjugate pair, it is only this solution, with $S_E>0$, that takes part in the saddle point expansion of the path integral. $z_c^*$ is discarded.

The classification of saddle points into relevant and irrelevant saddles is discussed in~\cite{Tanizaki:2014xba}. 
In a sentence, the restriction of the domain of functions that are included in the complexified path integral, is performed by defining a downward flow: a prescription that guarantees that all paths in the sum always posses a smaller value of ${\rm Re}\left(iS[z(t)]\right)$ than that achieved for the set of real-valued paths. Since real valued paths $r(t)$ always have a real-valued action, ${\rm Re}\left(iS[r(t)]\right)=0$, it follows that the functions participating in the complexified path integral must have ${\rm Re}\left(iS[z(t)]\right)<0$. For the tunneling problem, this maintains $z_c$ but removes $z_c^*$.

\subsection{Basic features of the tunneling solution}\label{ss:tunsol}
We now discuss basic features of the tunneling solution $z_c(t)$.

\begin{figure}[hbp!]
      \includegraphics[scale=0.42]{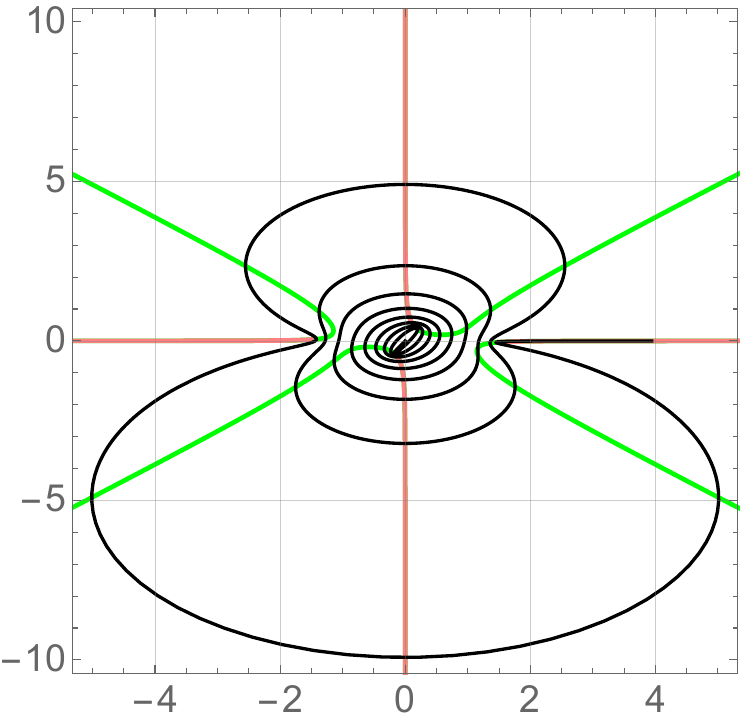} 
       \includegraphics[scale=0.45]{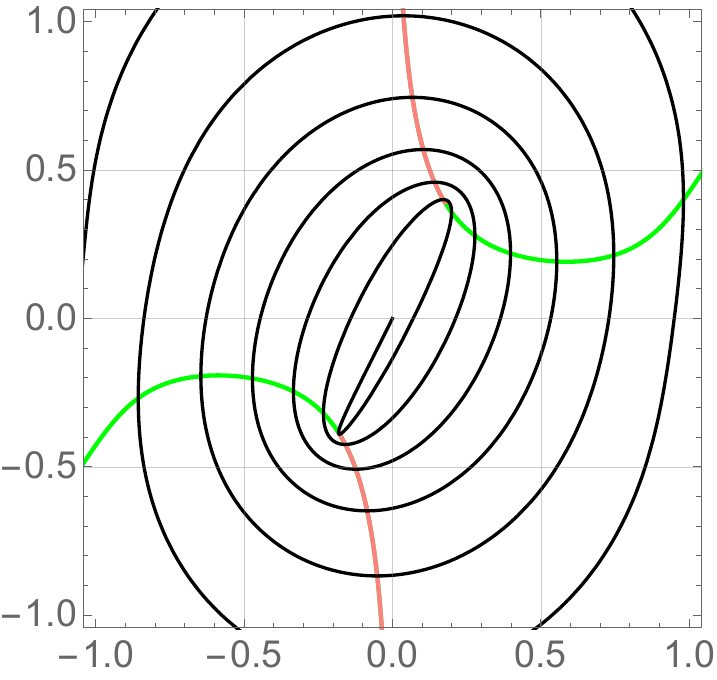} 
                  \includegraphics[scale=0.455]{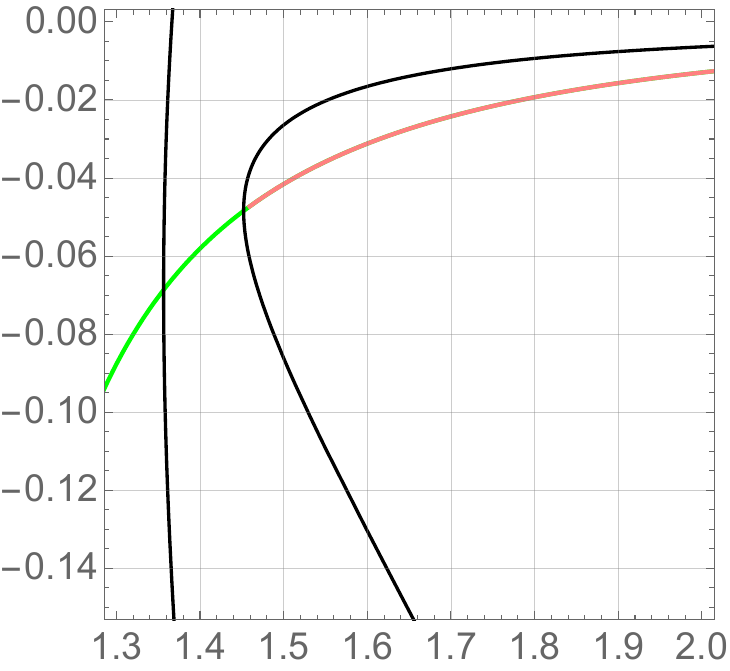} 
 \caption{{\bf Left panel:} tunneling solution ${z_c}(t)$ in the $({\rm Re}\,z,{\rm Im}\,z$) plane (black curving line). The solution starts at $z=0$ and emerges from the barrier region in the negative real $z$ direction. Along the green (orange) curves, the real (imaginary) part of ${\dot z_c}=\sqrt{2\epsilon-2V}$ vanishes. {\bf Middle panel:} zoom on the origin. {\bf Right panel:} zoom on the exit point. 
 }
 \label{fig:bounce1}
\end{figure}

For $n=4$, ${z_c}(t)$ can be found explicitly in terms of Jacobi functions~\cite{Turok:2013dfa}, and is characterized by\footnote{The solution we plot here is written in Mathematica by $z_c(t)=-\sqrt{\frac{2q}{1+q}}{\rm JacobiSN\left[\frac{t}{\sqrt{1+q}},q\right]}$. We note that Ref.~\cite{Turok:2013dfa} also studied the role of this solution for tunneling, with modified boundary condition at $t=0$ following the choice to employ the saddle point approximation on the wave function $\psi(y,t)=\int dx\,\psi_0(x)K(t;y,x)$, rather than $K(t;y,x)$.} $\epsilon=q/(1+q^2)$ with complex $q$. An example with $q=-0.06+0.06172i$ is shown in the {\bf left panel} of Fig.~\ref{fig:bounce1}. This solution starts at $x=0$ and escapes in the positive real $z$ direction. Along the green curves, the real part of ${\dot z_c}=\sqrt{2\epsilon-2V}$ vanishes; along the orange curves, the imaginary part of ${\dot z_c}$ vanishes; noting how the path changes direction in crossing these curves helps to understand some features of the solution. 

For small $\epsilon$, the path wraps multiple times around the points $z_\pm$ that provide the small-$z$ solutions of the equation
\be{\dot z_c}^2&=&2\epsilon-2V(z)=0.\ee
For our potential, these zero points satisfy $2\epsilon-z^2+\frac{2}{n}z^n=0$, and always contain a small-$z$ pair
\be\label{eq:zpm} z_\pm&=&\pm\sqrt{2\epsilon}\left(1+\frac{\left(\pm\sqrt{2\epsilon}\right)^{n-2}}{n}+...\right).\ee
We focus on the small-$z$ region in the {\bf middle panel} of Fig.~\ref{fig:bounce1}. The points $z_\pm$ can be identified on the plot as the contact points of the green and orange curves. 

Note that $z_\pm$, and in general, the set of points satisfying ${\dot z_c}^2=0$, are the branch points of the function ${\dot z_c}(z)=\sqrt{2\epsilon-2V(z)}$. A set of branch cuts can be constructed that extends ${\dot z_c}(z)$ to an analytic function in the complex $z$ plane. The branch cut associated to $z_\pm$ can be chosen as the line connecting $z_\pm$. The path in Fig.~\ref{fig:bounce1} circles both $z_\pm$ simultaneously, thereby wrapping around the cut without crossing it. The path does not wrap around the other branch points. This will become useful for us later.

\begin{figure}[h]
            \includegraphics[scale=0.6]{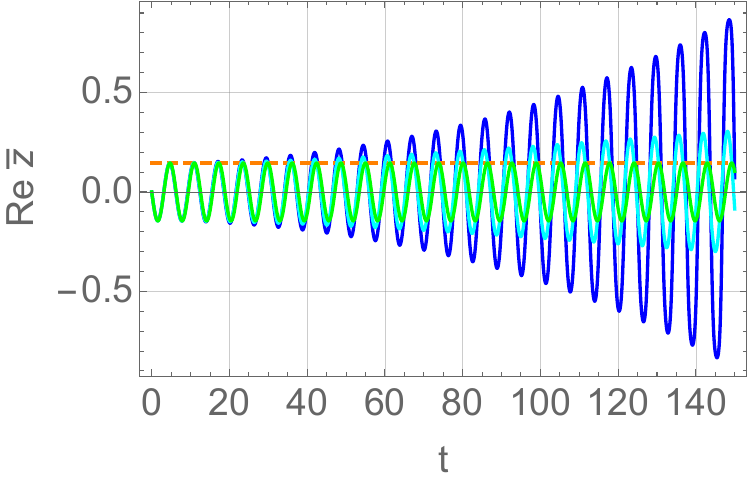} 
                              \includegraphics[scale=0.6]{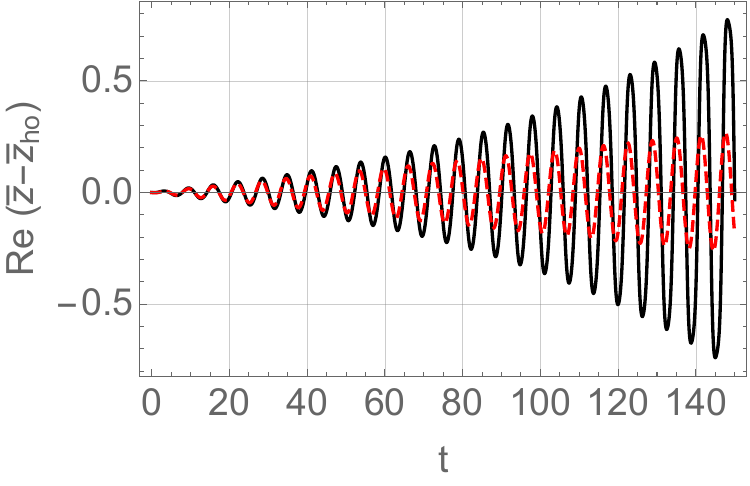} 
 \caption{{\bf Left panel:} In blue, we show ${\rm Re} {z_c}$ vs. $t$, illustrating the behavior at small $z$. In green we show ${\rm Re} {z_c}_{\rm ho}$, and in cyan ${\rm Re} ({z_c}_{\rm ho}+\dot a t\cos t)$, with constant $a$ taken from Eq.~(\ref{eq:a}). Dashed orange shows ${\rm Re}\sqrt{2|\epsilon|}$. {\bf Right panel:} In black we show ${\rm Re} ({z_c}-{z_c}_{\rm ho})$ vs. $t$. In dashed red we show $a t\cos t$.
 }
 \label{fig:drez}
\end{figure}

The density of the inner spiral increases as $\epsilon$ decreases (see~\cite{Turok:2013dfa,Cherman:2014sba} for related discussions), and the outward spiraling structure is driven by the nonlinear terms in $V(z)$. To see this, note that neglecting the nonlinear terms (that is, approximating $V(z)\approx\frac{1}{2}z^2$) we would have the harmonic oscillator solution ${z}_{\rm ho}(t)=x\cos t+s\sqrt{2\epsilon-x^2}\sin t$ with constant  $x$ and a sign choice $s=\pm1$, describing a closed ellipse (or a line if $x=0$) with period $2\pi$. We can set $x$ real and positive by choosing the phase of the cycle. The orbit goes counter-clockwise if ${\rm Im}\,s\sqrt{2\epsilon-x^2}>0$, and vice-verse. For a solution that starts at the FV minimum ($x^2\ll|2\epsilon|$), the sense of rotation is fixed by $s\,{\rm Im}\sqrt{\epsilon}$. 
%
To observe the slow outward drift of the spiral at small $z$, write $z_{c}(t)=z_{\rm ho}(t)+f(t)$, and solve for $f$ with the boundary condition $f(0)=0$. Letting $x\ll\sqrt{2|\epsilon|}$ to simplify the analysis, the EOM reads
${\ddot z_c}+V'
=f+\ddot f-\left({z}_{\rm ho}+f\right)^{n-1}\approx f+\ddot f-\left(\sqrt{2\epsilon}\sin t+f\right)^{n-1}=0$, 
so at leading order in $\epsilon$ we have $f\propto\left(\sqrt{2\epsilon}\right)^{n-1}$. It is straightforward to show that for {\bf even $n$}, $f$ contains a non-harmonic term $f\supset \left(\sqrt{2\epsilon}\right)^{n-1}t\,\cos t$, in addition to sine and cosine functions. The $\sim t\,\cos t$ term is responsible for the outward motion. For {\bf odd $n$}, the $t\,\cos t$ term arises only in the next order in the expansion, and one finds $f\supset \left(\sqrt{2\epsilon}\right)^{2n-3}t\,\cos t$. The result of this is that the full solution has the behavior $z_c=z_{\rm ho}+a\,t\,\cos t+...$, where $(...)$ contains harmonic functions with amplitude $\sim\left(\sqrt{2\epsilon}\right)^{n-1}$, and
\be\label{eq:a}a&\sim&\left\{\begin{array}{cc}\left(\sqrt{2\epsilon}\right)^{n-1},&{\rm n\;even}\\
\left(\sqrt{2\epsilon}\right)^{2n-3},&{\rm n\;odd}\end{array}\right\}.\ee 
We illustrate this behavior in Fig.~\ref{fig:drez}, for $n=4$, where $a=\frac{3}{2\sqrt{2}}\epsilon^{\frac{3}{2}}$.

Eq.~(\ref{eq:a}) shows that the modulus of out-spiraling solutions that start at $x\lesssim\sqrt{2|\epsilon|}$ stays in the vicinity of $\sqrt{2|\epsilon|}$ for a long duration of time, $t\sim\left|\sqrt{\epsilon}/a\right|\sim|\epsilon|^{1-\frac{n}{2}}$ for  $n$ even, or $t\sim\left|\sqrt{\epsilon}/a\right|\sim|\epsilon|^{2-n}$ for $n$ odd. This estimate misses logarithmic corrections, as we will discuss further below. 

The analysis leading to Eq.~(\ref{eq:a}) holds only as long as ${\rm Im}\,\epsilon^{\frac{n}{2}-1}\neq0$. Otherwise, one can show that no spiral motion develops: the orbit is trapped in the FV region and does not escape. We found it most convenient to explain this point using  analysis tools that we explain in the next section; we therefore defer the explanation to App.~\ref{s:app1}.

The path in Fig.~\ref{fig:bounce1} exits the barrier along positive real $z$. We highlight the exit point in the {\bf right panel}. After escaping, the solution is sandwiched between the real line and the curve ${\rm Im}\,{\dot z_c}=0$. At large positive $z$ near the real axis, $|z_r|>1\gg |z_i|$ (with the notation $z=z_r+iz_i$, and similarly for $\epsilon$), we have ${\rm Im}\,{\dot z_c}\approx{\rm Im}\sqrt{2\epsilon_r+\frac{2}{n}z_r^{2n}+2i(\epsilon_i+3z_rz_i)}$, 
so ${\rm Im}\,{\dot z_c}=0$ is described by $z_i\approx-\epsilon_i/(3z_r)$. Therefore, the solution approaches a real endpoint $y$ as ${\rm Im}\,z\propto1/y$. 

\subsection{Action calculation}
We now show that $S[{z_c}]$ reduces to a simple generic expression when $\epsilon$ is small. At leading order in $\epsilon$, there is no need to find ${z_c}(t)$ explicitly, in order to calculate the action. This conclusion extends also to the fluctuation integral. 

Rewriting the action as a contour integral\footnote{Ref.~\cite{Behtash:2015loa} discussed a related analysis, but there, the complexification is done for the path integral of the Euclidean theory, with solutions calculating the spectrum of states trapped in the potential, rather than the tunneling wave function.} along ${z_c}$, via $dt=dz/{\dot z_c}$, and using $\frac{1}{2}{\dot z_c}^2-V=2\epsilon-2V-\epsilon$, we can write
\be\label{eq:Szb1} S[{z_c}]&=&\int_{\mathrm{{z_c}}}dz\sqrt{2\epsilon-2V(z)}-\epsilon t,
\ee
with
\be\label{eq:tint}
t&=&\int_{\mathrm{{z_c}}}\frac{dz}{{\dot z_c}}=\int_{\mathrm{{z_c}}}\frac{dz}{\sqrt{2\epsilon-2V(z)}}.\ee
We should make a couple of comments about the passage from time integral to contour integral along $z_c$. First, note that the branch cuts of $\dot z_c$ can be constructed such that the tunneling path never crosses a cut. Of course, the limit $\epsilon\to0$ needs to be handled with care, because in this limit the path comes arbitrarily close to branch points. Second, the tunneling path of least action is not periodic (in $t$), so it defines a proper one-to-one map $t=t(z_c)$. There are, in general, tunneling paths that fold back upon themselves to include near exact cycles; for example, a slight deformation of $\epsilon$ could reflect $z_c$ back on its tracks when it hits the classical turning point $z\approx b$. But such paths have an exponentially suppressed action in comparison to the ``primary" $z_c$ we focus on, and consequently, we do not concern ourselves about them. (In the imaginary time analysis, these are multi-bounce configurations.)

Consider the integral $\int_{\mathrm{{z_c}}}dz\sqrt{2\epsilon-2V(z)}$ in Eq.~(\ref{eq:Szb1}). 
The starting point of the path is ${z_c}(0)=x$ and the end point is ${z_c}(t)=y$. We can add and subtract an integral from $x$ to $y$ along the real $z$ axis; 
this splits ${z_c}$ into a closed contour starting and ending at $z=x$ (call this contour $\mathcal{C}_{{z_c}}$), plus the integral on the real line:
\be\label{eq:mainint} \int_{\mathrm{{z_c}}}dz\sqrt{2\epsilon-2V(z)}&=&\mathcal{I}_{{z_c}}+\mathcal{S},\\
\label{eq:curlS}\mathcal{S}&=&\int_x^y dr\sqrt{2\epsilon-2V(r)},\\
\label{eq:Izb}\mathcal{I}_{{z_c}}&=&\oint_{\mathcal{C}_{{z_c}}}dz\sqrt{2\epsilon-2V(z)}.
\ee

\begin{figure}[h]
            \includegraphics[scale=0.475]{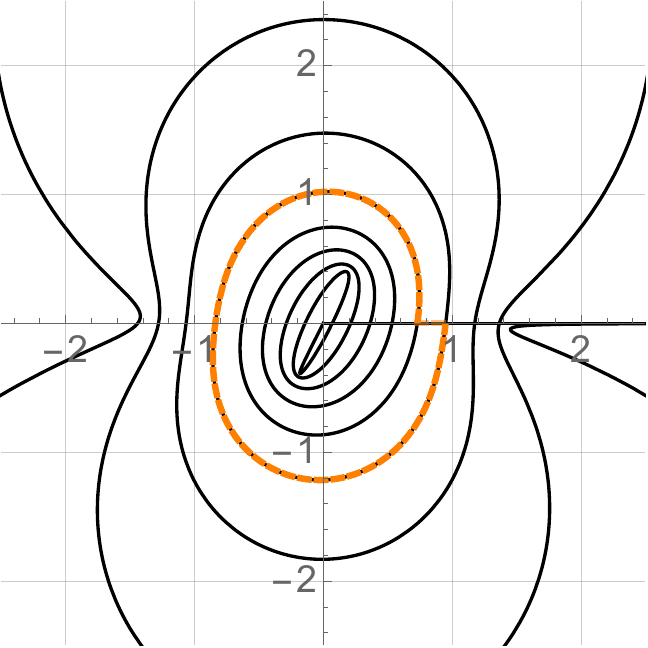} 
 \caption{$\mathcal{C}_{{z_c}}$ with one cycle highlighted. 
 }
 \label{fig:loop}
\end{figure}

Let us first analyze $\mathcal{I}_{{z_c}}$. Recall that ${z_c}$ wraps multiple times around the pair of zeros $z_\pm$ of ${\dot z_c}$ (see Eq.~(\ref{eq:zpm})). $\mathcal{I}_{{z_c}}$ can therefore be evaluated as ${N}$ closed cycles around $z_\pm$, where ${N}$ is a positive integer or half-integer (${N}\gg1$ if $t\gg1$, so ${N}+0.5$ can be approximated as ${N}$). This is illustrated in Fig.~\ref{fig:loop}.
As long as we do not cross branch cuts, we can deform the contour of integration. Thus, the $N$ cycles all give the same result. Each cycle can be shrunk around $z_\pm$; denote the single small cycle $\mathcal{C}^1_\epsilon$:
\be \mathcal{I}_{{z_c}}&=&{N}\oint_{\mathcal{C}^1_\epsilon}dz\sqrt{2\epsilon-2V(z)}.\ee
Considering the result as an expansion in powers of $\epsilon$, we can expand the potential to leading order, and then map to the unit circle via $z=\sqrt{2\epsilon}\zeta$, noting that $|z_\pm|=\sqrt{2|\epsilon|}$ at leading order in $\epsilon$. With this,
\be\label{eq:Ceps}\oint_{\mathcal{C}^1_\epsilon}dz\sqrt{2\epsilon-z^2}&\approx&2\epsilon\oint_{\mathcal{C}^1_1}d\zeta\sqrt{1-\zeta^2}=2\pi\epsilon,\ee
where $\mathcal{C}^1_1$ is the unit circle. Here, we evaluated the integral as two equal contributions, one for each side of the $(z_-,z_+)$ cut: $\oint_{\mathcal{C}^1_1}d\zeta\sqrt{1-\zeta^2}=-2\int_0^\pi d\phi\sqrt{1-e^{2i\phi}}\left(-\sin\phi+i\cos\phi\right)=\pi$. This exercise of extending the square-root on both sides of the cut is equivalent to the physical requirement that the velocity of the trajectory be continuous\footnote{The over-all sign on the RHS of Eq.~(\ref{eq:Ceps}) is somewhat tricky. Obtaining it requires matching $dz$ to ${\dot z_c}$ along ${z_c}$, and noting that the point $\zeta=e^{i0_+}$, from where we start the unit cycle integration, is located just across the cut.}. 

We are going to need the leading correction to Eq.~(\ref{eq:Ceps}) at the next order in $\epsilon$. Unlike the leading $2\pi\epsilon$ term, the correction depends on the details of the nonlinear interactions in $V$. We can arrange the correction as two contributions: one coming directly from the nonlinear term $\sim z^n$ in $V$, and another coming from the $\epsilon$ corrections to $z_\pm$ in Eq.~(\ref{eq:zpm}). The latter effect is captured in the mapping of $\mathcal{C}^1_\epsilon$ to $\mathcal{C}^1_1$, which should now read $z=\sqrt{2\epsilon}\left(1+\frac{1}{n}\left(\sqrt{\pm2\epsilon}\right)^{n-2}+...\right)\zeta$, where $(...)$ denotes terms at higher orders in $\epsilon$. To simplify the derivation, let us assume that $n$ is even -- it is easy to generalize the result later on. For even $n$, we have
\be \oint_{\mathcal{C}^1_\epsilon}dz\sqrt{2\epsilon-z^2+\frac{2z^n}{n}}
&=&2\epsilon\left(1+\frac{1}{n}\left(\sqrt{2\epsilon}\right)^{n-2}\right)\oint_{\mathcal{C}^1_1}d\zeta\sqrt{1-\zeta^2+\frac{2}{n}\left(\sqrt{2\epsilon}\right)^{n-2}\left(\zeta^n-\zeta^2\right)}+...\no\\
\label{eq:inteps}&=&2\pi\epsilon\left(1+\frac{2\Gamma\left(\frac{n}{2}+\frac{1}{2}\right)}{n\sqrt{\pi}\Gamma\left(\frac{n}{2}+1\right)}\left(\sqrt{2\epsilon}\right)^{n-2}+...\right),\;\;\;\;\;\;\;\;({\rm for\;even} \;n)
\ee
For odd $n$, one finds that the correction term in the brackets begins at $\mathcal{O}\left(\epsilon^{n-2}\right)$, instead of the $\mathcal{O}\left(\epsilon^{\frac{n}{2}-1}\right)$ of Eq.~(\ref{eq:inteps}). We conclude that
\be\label{eq:Izscale}\mathcal{I}_{{z_c}}&=&2\pi{N}\epsilon\left(1+A_{\mathcal{I}}\epsilon^{\frac{n}{2}-1}+...\right),\ee
where the numerical coefficient $A_{\mathcal{I}}$ depends on the details of the nonlinear terms in $V$.

Next, we analyze the $\epsilon$-dependence of $\mathcal{S}$ (Eq.~(\ref{eq:curlS})). It is clear that $\mathcal{S}$ has a finite nonzero limit as $\epsilon\to0$; this limiting value will be seen to relate to Coleman's Euclidean action. What we are after, however, is a non-analytic piece $\mathcal{S}\supset\epsilon\ln\epsilon$. Identifying this term will become useful to clarifying the connection between $\epsilon$ and $t$. 

In thinking about this integral, it is useful to define the distance scale $r_{nl}$, below which nonlinear interactions in $V$ become unimportant in ${\dot z_c}$:
\be r_{nl}&=&\left(n|\epsilon|\right)^{\frac{1}{n}}.\ee
We defined $r_{nl}$ such that at $r<r_{nl}$, the term $\frac{2}{n}r^n$ in $2V(r)$ contributes less than $2|\epsilon|$. 
Note that for $n\geq3$ we have $r_{nl}\gg\sqrt{2|\epsilon|}$ at small $\epsilon$. Therefore in general we can find some reference scale, call it $l$, that satisfies $\sqrt{2|\epsilon|}\ll l\ll r_{nl}$. 
The $\mathcal{S}$ integral can then be written as two parts, $r<l$ and $r>l$. In the former, nonlinear terms can be omitted:
\be\mathcal{S}
&\approx&\int_x^ldr\sqrt{2\epsilon-r^2}+\int_l^ydr\sqrt{2\epsilon-2V(r)}.\ee
Because the starting point $x$ of ${z_c}$ enters the calculation of $\mathcal{S}$ explicitly, we tentatively allowed $x\neq0$. We still think of $x$ as small, specifically $x\ll r_{nl}$. 

The $r<l$ integral gives
\be\int_x^ldr\sqrt{2\epsilon-r^2}
&=&i\frac{l^2}{2}-\frac{x}{2}\sqrt{2\epsilon-x^2}+\frac{i\epsilon}{2}\left[\ln\left(\frac{-\frac{\epsilon^2}{2l^2}}{\epsilon-x^2-ix\sqrt{2\epsilon-x^2}}\right)-1\right].\ee
For $x\ll\sqrt{2|\epsilon|}$, we can neglect $x$ to obtain: 
\be\int_x^ldr\sqrt{2\epsilon-r^2}&=&\frac{i}{2}\left[\epsilon\ln\epsilon+\epsilon\left(i\pi-\ln(2l^2)-1\right)+l^2+...\right],\,\,\,\,\,\,\left({\rm for}\;\;x\ll\sqrt{2|\epsilon|}\right).\ee
This contains the $\epsilon\ln\epsilon$ term we were looking for. This term is always present if $x$ is very small, specifically if $x\to0$ as in Coleman's original analysis. For $x\gg\sqrt{2|\epsilon|}$, the explicit $\epsilon\ln\epsilon$ term is not there. Instead we have
\be\int_x^ldr\sqrt{2\epsilon-r^2}&=&i\frac{l^2-x^2}{2}+i\epsilon\ln\frac{x}{l}+...,\,\,\,\,\,\,\left({\rm for}\;\;x\gg\sqrt{2|\epsilon|}\right).\ee
We see that the logarithmic enhancement goes away for $x\gg\sqrt{2|\epsilon|}$, and disappears for $x\sim r_{nl}$.

Considering the $r>l$ integral, it is not difficult to see that this does not give an $\epsilon\ln\epsilon$ term: only a constant (in $\epsilon$) plus $\mathcal{O}(\epsilon)$ terms. The details of these terms will not be needed for us. 

Altogether we conclude that the $\epsilon$ scaling of $\mathcal{S}$ is
\be\label{eq:Sscale}\mathcal{S}&=&\mathcal{S}_{0}+iA_{\mathcal{S}}\epsilon\ln\epsilon+B_{\mathcal{S}}\epsilon+...\,,\ee
where
\be\label{eq:S0}\mathcal{S}_{0}&=&\int_x^ydr\sqrt{-2V(r)}.\ee
For small $x\ll\sqrt{2|\epsilon|}$, the coefficient $A_{\mathcal{S}}=\frac{1}{2}$ is independent of the details of $V$.

Before we return to calculating the action, we pause to derive a useful relation. 
From Eqs.~(\ref{eq:tint}),~(\ref{eq:Izscale}), and~(\ref{eq:Sscale}), we find\footnote{$\mathcal{I}_{z_c}+\mathcal{S}$ is not an analytic function of $\epsilon$. Indeed, $\mathcal{S}\supset\epsilon\ln\epsilon$, and $\mathcal{I}_{z_c}$ contains a simple pole. Expressing $t$ as an $\epsilon$-derivative of this function may seem awkward, but this shortcut expression works because we are interested precisely in extracting the leading divergence of the derivative as $\epsilon$ becomes small (but never truly zero).}:
\be\label{eq:tde} t&=&\frac{d}{d\epsilon}\left(\mathcal{I}_{{z_c}}+\mathcal{S}\right)\\
&=&2\pi{N}\left(1+\frac{n}{2}A_{\mathcal{I}}\epsilon^{\frac{n}{2}-1}\right)+iA_{\mathcal{S}}\ln\epsilon+iA_{\mathcal{S}}+B_{\mathcal{S}}+...\,.\ee
Imposing ${\rm Im}\,t=0$, we have
\be\label{eq:Nepsscale}{N}&=&\frac{A_{\mathcal{S}}}{n\pi A_{\mathcal{I}}}\frac{-\ln|\epsilon|}{{\rm Im}\epsilon^{\frac{n}{2}-1}}+...\,.\ee
We have already seen, from the analysis of the inner spiraling structure of ${z_c}$ (Sec.~\ref{ss:tunsol}), that ${N}\sim\epsilon^{1-\frac{n}{2}}$ if the starting point $x$ is close to the FV minimum. Eq.~(\ref{eq:Nepsscale}) sharpens this result including log corrections.

Finally, we turn to the action. Using Eqs.~(\ref{eq:Izscale}),~(\ref{eq:Sscale}), and~(\ref{eq:Nepsscale}), we find
\be\label{eq:Szbfinal} S[{z_c}]&=&\left(1-\epsilon\frac{d}{d\epsilon}\right)\left(\mathcal{I}_{{z_c}}+\mathcal{S}\right)\no\\
&=&\mathcal{S}_0-iA_{\mathcal{S}}\epsilon-2\pi\left(\frac{n}{2}-1\right)A_{\mathcal{I}}{N}\epsilon^{\frac{n}{2}}\no\\
&=&\mathcal{S}_0+\mathcal{O}\left(\epsilon\ln\epsilon\right).\ee
In the last line, we have assumed that the real part of $\epsilon$ is not parametrically large compared with the imaginary part\footnote{We did not find a simple argument to justify this assumption. Indeed, we suspect that saddle point solutions with small $|\epsilon|$ but large ${\rm Re}\,\epsilon/{\rm Im}\,\epsilon$ exist, related to the decay of excited states of the FV region. Nevertheless, we also expect that the decay of states near to the ground level of the FV region does not exhibit this kind of hierarchy, and it would be such solutions that dominate the large $t$ wave function.}.

Inspecting $\mathcal{S}_0$ (Eq.~(\ref{eq:S0})), we can summarize that the small-$\epsilon$ saddle point contribution to the propagator is constructed from a part that produces the usual exponential suppression, coming from the integral between the starting point $x$ and the classical turning point $b$; and a pure phase part corresponding to the action of a free particle rolling with zero energy from $b$ to the endpoint $y$:
\be\label{eq:Ksmalleps} iS[{z_c}]&\approx&-S_E+iS_{\rm free},\\
\label{eq:SE}S_E&=&\int_x^bdr\sqrt{2V(r)},\\
\label{eq:Sfree}S_{\rm free}&=&\int _b^ydr\sqrt{-2V(r)}.\ee
Both $S_E$ and $S_{\rm free}$ are real and positive. $S_E$ is, of course, a generalization of Coleman's Euclidean action~\cite{Coleman:1977py}. Altogether, Eq.~(\ref{eq:Ksmalleps}) coincides with the usual WKB expression for the wave function. This is the bottom line we were getting at, and  concludes our exercise of unraveling the bounce.

\section{Discussion}\label{s:discuss}
We now discuss a few aspects of our derivation.

\begin{enumerate}
\item {\bf Decay law.} Eq.~(\ref{eq:Ksmalleps}) reproduces the exponential decay law. 
A quick (and standard) way to see this, is to extend the FV probability $P_{\rm FV}$ (see Sec.~\ref{s:int}) to the interval $(-\infty,y)$, namely, count the total probability to find the particle to the left of $y$. Call this 
\be P_{<y}&=&\int_{-\infty}^ydx|\psi(x,t)|^2.\ee
In terms of the probability current $j(x,t)={\rm Im}\,\psi^*\partial_x\psi$,  
we have 
\be \dot P_{<y}&=&-j(y,t),\ee
so as long as $P_{<y}(t)$ is of order unity, we can extract the decay rate from $\Gamma=j$.  
Using our result for the semiclassical propagator, the wave function is given by
\be\psi(y,t)&=&\left[\int dx\psi_0(x)\mathcal{A}e^{-S_E}\right]e^{iS_{\rm free}}.\ee
To directly compare our results to Coleman's formalism, we can let $\psi_0(x)\approx\delta(x)$; in that case $\int dx\psi_0(x)\mathcal{A}e^{-S_E}\approx \mathcal{A}e^{-S_E}$ evaluated at\footnote{More generally, we would have $j(y,t)={\rm Im}\int dx'\int dx\,\psi^*_0(x')\psi_0(x)e^{-S_E(x)-S_E(x')}\mathcal{A}_{x'y}^*\mathcal{A}_{xy}\left(\frac{\partial_y\mathcal{A}_{xy}}{\mathcal{A}_{xy}}+ip(y)\right)$, where we manifest explicitly the $x$ dependence of $S_E$ from Eq.~(\ref{eq:SE}), and the $x$ and $y$ dependence of $\mathcal{A}$.} $x=0$. 
Neglecting the $y$ dependence of the fluctuation term $\mathcal{A}$ compared with the exponential, we find 
\be j(y,t)&\approx&2\left|\mathcal{A}\right|^2p(y)e^{-2S_E},\;\;\;\;\;p(y)=\partial_yS_{\rm free}=\sqrt{-2V(y)}.\ee
Note that $p(y)$ is the classical momentum of a zero energy particle rolling down the potential from $b$ to $y$. We see $\Gamma= j\approx2\left|\mathcal{A}\right|^2p\,e^{-2S_E}$.

We have not calculated the fluctuation prefactor $\mathcal{A}$. From the standard (Schr\"odinger-based) WKB analysis, we can expect $|\mathcal{A}(y)|^2\propto 1/\sqrt{-2V(y)}=1/\sqrt{p(y)}$. This scaling is associated with current conservation: the flux crossing any point $y$ must be independent of $y$ at large $t$.

Neglecting $\partial_y\mathcal{A}\sim\frac{\partial_yp}{p}\mathcal{A}$ as compared to $\mathcal{A}\partial_yS_{\rm free}\sim p\mathcal{A}$ is also a standard step in the semiclassical approximation. This does not mean that this step is generally valid. Its validity boils down to the requirement $p^2\gg\partial_yp$: for our potential $p(y)\sim y^{\frac{n}{2}}$ outside of the barrier, so the approximation is valid at sufficiently large $y$. Near the exit point, $y\sim b$, we have $p\to0$ while $\partial_yp\neq0$, and the WKB approximation is potentially poor. This reinforces the intuition that the semiclassical approximation falls apart at the edge of the barrier, but improves as the particles roll further downhill becoming more and more ``classical"~\cite{Guth:1985ya}.

\item {\bf Exploring $x\neq0$ in Eq.~(\ref{eq:SE}).} The essential parts of our analysis hold also if we allow $x\neq0$ in Eq.~(\ref{eq:SE}), at least for small $|x|\sim\sqrt{2|\epsilon|}$. Importantly, the lower integration limit of Eq.~(\ref{eq:SE}) combines with initial data $\psi_0(x)\sim e^{-\frac{x^2}{2}}$  (namely, with the wave function of the would-be ground state of the FV) to give $\psi_0(x)e^{-\int_x^bdr\sqrt{2V(r)}}\approx \psi_0(0)e^{-\int_0^bdr\sqrt{2V(r)}}$, independent of $x$. This is of course not an accident: the would-be FV ground state wave function can be estimated with a small twist on the real time analysis we reported (or equally well, in the standard imaginary time technique) to scale as $\psi_0(x)\sim e^{-\int_0^xdr\sqrt{2V(r)}}$, which precisely complements the tunneling term $e^{-\int_x^bdr\sqrt{2V(r)}}$ in the propagator. Neglecting the $x$-dependence of $\mathcal{A}$ (which by symmetry reasons, must have vanishing first derivative at $x=0$, exactly for even $n$, and at lowest order in the interactions for any $n$), this gives
\be\int dx\psi_0(x)\mathcal{A}e^{-\int_x^bdr\sqrt{2V(r)}}&\sim&\left(\psi_0\mathcal{A}e^{-S_E}\right)_{x=0}.\ee

\item{\bf The bounce.} Coleman's bounce~\cite{Coleman:1977py} can be used to parameterize Eq.~(\ref{eq:SE}) by defining  ``imaginary time" $\tau(r)$ via $\tau(r)=\int_r^b\frac{dr'}{\sqrt{2V(r')}}$ for $r$ in the range $(0,b)$. The inverted function $r(\tau)$, monotonically decreasing from $r=b$ at $\tau=0$ to $r\to0$ at $\tau\to\infty$, satisfies Coleman's bounce equations $\frac{1}{2}\dot r^2-V(r)=0$ and $\ddot r-\partial_r V=0$. Eq.~(\ref{eq:SE}) (with $x\to0$) becomes
\be\label{eq:Eucbounce}S_E&=&\int_0^bdr\sqrt{2V(r)}=\int_0^\infty d\tau\left(\frac{1}{2}\dot r^2+V(r)\right).\ee

The calculation we did is not Coleman's FV-to-FV calculation, but FV-to-free region. However, our analysis carries to Coleman's if we let $y\to x\to0$ and note that the lowest $\epsilon$ real time solution resembles the path in the {\bf left panel} of Fig.~\ref{fig:bounce1}, apart from that instead of monotonously out-spiraling towards the exit point $b$, it strikes $b$ earlier on, and inspirals back to the origin. This solution has action $2S_E$, up to finite-$\epsilon$ corrections as we calculated.  


\item {\bf WKB-like factorization, $\epsilon$ corrections.} The factorization of the propagator into an ``imaginary time" piece and a "real time" piece was discussed in~\cite{Bramberger:2016yog}. Of course, this factorization was also expected from the usual WKB calculation. Our derivation, apart from providing a somewhat different pedagogical perspective, may add to this analysis the ability to incorporate finite-time corrections via higher-order $\epsilon$ terms. For the simple unbounded polynomial potential we considered, Eq.~(\ref{eq:Szbfinal}) shows that
\be\label{eq:Szcepscorr} S[z_c]&=&\mathcal{S}_0+\left(\frac{1}{2}-\frac{1}{n}\right)\frac{\epsilon^{\frac{n}{2}}\ln|\epsilon|}{{\rm Im}\epsilon^{\frac{n}{2}-1}}+{\rm higher\;powers\;of\;}\epsilon.\ee

\item {\bf Multi-instanton configurations.} 
It is natural to guess that multi-instanton solutions extend the ``fundamental" solution we analyzed.  These solutions would resemble the path in the {\bf left panel} of Fig.~\ref{fig:bounce1}, but recoil back and forth between $b$ and the origin before finally exiting. In the small $\epsilon$ limit, $m$ back-and-forth detours before final exit would contribute a factor of $e^{-2mS_E}$ to the propagator, the usual multi-instanton suppression.

If this picture is correct, then the $\epsilon$ expansion may help to test the validity of the $e^{-2mS_E}$ approximation. An $m$-instanton configuration must still make it to the final destination $y$ by time $t$. Since each single cycle around the FV region lasts $\Delta t\approx2\pi$, the total number of cycles $N$ must be the same as for the $m=0$ ``fundamental" solution. This means that an $m$-instanton path out-spirals from $x\approx0$ to $x\approx b$ in $N/(2m+1)$ cycles. This leads to a modified version of Eq.~(\ref{eq:Nepsscale}): $N\approx(2m+1)\frac{A_{\mathcal{S}}}{n\pi A_{\mathcal{I}}}\frac{-\ln|\epsilon_m|}{{\rm Im}\epsilon_m^{\frac{n}{2}-1}}$. Inverting this relation shows that the $\epsilon_m$ of $m$-configurations is larger than the $\epsilon$ of the fundamental solution. For example, considering the quartic potential $n=4$, we expect $\epsilon_m\sim(2m+1)\epsilon$, up to log corrections. Referring back to Eq.~(\ref{eq:Szcepscorr}), we expect that finite-$\epsilon$ corrections  start to clatter the imaginary time limit for sufficiently high-order (high $m$) multi-instanton configurations.

\item {\bf Finite-time expansion.} $\epsilon$-corrections map to finite-time corrections via $t\approx2\pi N$ and Eq.~(\ref{eq:Nepsscale}); e.g., for $n=4$, $\mathcal{O}\left(\frac{\ln\epsilon}{\epsilon}\right)=\mathcal{O}\left(t\right)$ (see~\cite{Turok:2013dfa} for closely related discussion of this point, including a consistent derivation of the time--energy relation). Up to the possibility of parametric hierarchy between ${\rm Re}\,\epsilon$ and ${\rm Im}\,\epsilon$ (a hierarchy that -- we should note -- we were not completely able to exclude), this could allow one to organize the analysis of multi-instanton corrections in terms of a $1/t$ expansion (see Ref.~\cite{Pimentel:2019otp} for related discussion).

\end{enumerate}

\section{Summary}
We presented an analysis of the saddle point tunneling solution of the complexified classical equations of motion (EOM), that dominates the wave function at large times when calculated using the real time path integral. Our goal was to examine how this saddle point unravels to give Coleman's  imaginary time result; or, similarly, the Schr\"odinger-based stationary WKB result; while keeping tabs on finite-time corrections. We did this exercise by organizing the calculation in powers of the energy $\epsilon$ characterizing the path.
Our analysis differs from previous literature in that we track the real time analytically-continued complex path, rather than performing the analytic continuation w.r.t. to the time variable itself. 

Apart from some pedagogical value (we think!), our derivation may also be useful for the analysis of finite-time corrections to the tunneling wave function. For example, although we did not explore this in detail, our approach may help to study the breakdown of naive multi-instanton resummation.

Extending our analysis to the fluctuation determinant is left for future work. At the time of writing, similar arguments to those presented above seem to successfully identify the usual imaginary time fluctuation integral at lowest order in the $\epsilon$ expansion, extending it in a natural way out to the free region of the potential. However, we are still bogged down by some questions related to the analyticity properties of complexified fluctuations. 

\acknowledgments
We thank Ofer Aharony, Shimon Levit, Ohad Mamroud, Mehrdad Mirbabayi, Yossi Nir, Gui Pimentel, Adam Schwimmer, Amit Sever, and Giovanni Villadoro for useful discussions. 
This work was supported by the Israel Science Foundation grant 1784/20, and by MINERVA grant 714123. 

\begin{appendix}
\section{Constraint on the phase of $\epsilon$ required for out-spiraling structure of $z_c$}\label{s:app1}
Let us calculate the time $\Delta t_1$ it takes $z_c$ to propagate from one crossing of the real $z$ axis, say at $z=r$, to the next crossing, $z=r+\Delta r$ (see Fig.~\ref{fig:loop}). As in the main text, we can write $\Delta t_1$ as the sum of a closed loop integral (going through the entire marked cycle in Fig.~\ref{fig:loop}), plus a short integral of length $\Delta r$ along the real $z$ axis. With an analysis similar to the main text, we readily obtain, for even $n$:
\be\label{eq:app1}\Delta t_1&=&\int _r^{r+\Delta r}\frac{dr}{\sqrt{2\epsilon-2V(r)}}+\oint_{\mathcal{C}^1_\epsilon}\frac{dz}{\sqrt{2\epsilon-2V(z)}}\\
&=&2\pi+n\pi A_{\mathcal{I}}{\rm Re}\,\epsilon^{\frac{n}{2}-1}-\frac{1}{2}{\rm Arg}\left(\frac{\epsilon-(r+\Delta r)^2-i(r+\Delta r)\sqrt{2\epsilon-(r+\Delta r)^2}}{\epsilon-r^2-ir\sqrt{2\epsilon-r^2}}\right)\no\\
&+&i\left[n\pi A_{\mathcal{I}}{\rm Im}\,\epsilon^{\frac{n}{2}-1}+\frac{1}{2}\ln\left|\frac{\epsilon-(r+\Delta r)^2-i(r+\Delta r)\sqrt{2\epsilon-(r+\Delta r)^2}}{\epsilon-r^2-ir\sqrt{2\epsilon-r^2}}\right|\right].\no\ee

Now, what we are calculating here is real time across the motion, so ${\rm Im}\,\Delta t_1=0$ must hold. This says that if ${\rm Im}\,\epsilon^{\frac{n}{2}-1}=0$, then the $\ln|...|$ term in the last line of Eq.~(\ref{eq:app1}) must vanish, namely, we must have $\Delta r=0$. Thus, for ${\rm Im}\,\epsilon^{\frac{n}{2}-1}=0$ the path must close-in on itself whenever it completes a cycle, and there cannot be any outward motion.

The analysis of the odd $n$ case is very similar, and leads to the same constraint: ${\rm Im}\,\epsilon^{\frac{n}{2}-1}\neq0$ is needed for outward motion.

\end{appendix}

\bibliography{ref}
\bibliographystyle{utphys}

\end{document}